\documentclass[journal=jctcce]{achemso}
\setkeys{acs}{articletitle = true}

\usepackage{graphicx,dcolumn,bm,xcolor,multirow,amsmath,amssymb,amsfonts,physics}
\usepackage[normalem]{ulem}
\usepackage[version=4]{mhchem}
\usepackage[compat=1.1.0]{tikz-feynman}
\usepackage[tracking]{microtype}
\usepackage{mathpazo,libertine}
\usepackage[
	colorlinks=true,
    citecolor=blue,
    breaklinks=true
	]{hyperref}
\definecolor{mypink}{rgb}{0.858, 0.188, 0.478}

\DeclareMathOperator{\sign}{sign}

\newcommand{\br}{\boldsymbol{r}}

\newcommand{\HF}{\text{HF}}	
	
\newcommand{\qsGW}{\text{qs$GW$}}	
\newcommand{\GOWO}{$G_0W_0$}	
		
\newcommand{\COHSEX}{\text{COHSEX}}

\newcommand{\Om}[1]{\Omega_{#1}}

\newcommand{\bOm}{\boldsymbol{\Omega}}
\newcommand{\bA}{\mathrm{\bf{A}}}
\newcommand{\bB}{\mathrm{\bf{B}}}
\newcommand{\bX}{\mathrm{\bf{X}}}
\newcommand{\bY}{\mathrm{\bf{Y}}}

\newcommand{\lcpq}{Laboratoire de Chimie et Physique Quantiques, Universit\'e de Toulouse, CNRS, UPS, France}
\newcommand{\lpt}{Laboratoire de Physique Th\'eorique, Universit\'e de Toulouse, CNRS, UPS, France}
\newcommand{\etsf}{European Theoretical Spectroscopy Facility (ETSF)}

\title{Potential energy surfaces without unphysical discontinuities: the Coulomb-hole plus screened exchange approach}

\author{J.~Arjan Berger}
\affiliation{\lcpq}
\altaffiliation{\etsf}
\email{arjan.berger@irsamc.ups-tlse.fr}
\author{Pierre-Fran\c{c}ois Loos}
\affiliation{\lcpq}
\author{Pina Romaniello}
\affiliation{\lpt}
\altaffiliation{\etsf}

\begin{document}	

\begin{tocentry}
\includegraphics{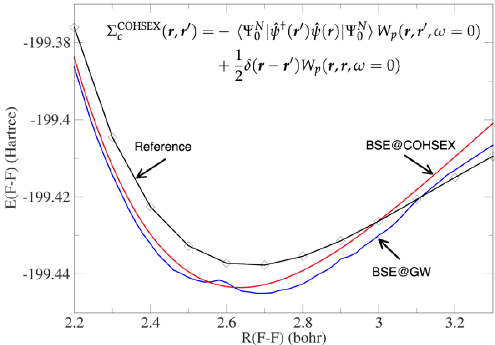}
\end{tocentry}

\begin{abstract}
In this work we show the advantages of using the Coulomb-hole plus screened-exchange (COHSEX) approach in the calculation of potential energy surfaces.
In particular, we demonstrate that, unlike perturbative $GW$ and partial self-consistent $GW$ approaches, such as eigenvalue-self-consistent $GW$ and quasi-particle self-consistent $GW$, the COHSEX approach yields smooth potential energy surfaces without irregularities and discontinuities.
Moreover, we show that the ground-state potential energy surfaces (PES) obtained from the Bethe-Salpeter equation, within the adiabatic connection fluctuation dissipation theorem, built with quasi-particle energies obtained from perturbative COHSEX on top of Hartree-Fock (BSE@COHSEX@HF) yield very accurate results for diatomic molecules close to their equilibrium distance.
When self-consistent COHSEX quasi-particle energies and orbitals are used to build the BSE equation the results become independent of the starting point.
We show that self-consistency worsens the total energies but improves the equilibrium distances with respect to BSE@COHSEX@HF.
This is mainly due to changes in the screening inside the BSE.
\end{abstract}

\maketitle

\section{Introduction}
In the last decade the $GW$ method \cite{Hedin_1965,Aryasetiawan_1998,Reining_2017,Golze_2019} 
has become a standard tool in the quantum-chemistry tool box.
It has proved to be a powerful approach for the calculation of ionization energies, electron affinities, fundamental gaps, etc.
However, due to the complexity of the $GW$ self-energy which is non-Hermitian and frequency dependent, a fully self-consistent approach is nontrivial. \cite{Stan_2006, Stan_2009, Rostgaard_2010, Caruso_2012, Caruso_2013, Caruso_2013a, Caruso_2013b, Koval_2014, Wilhelm_2018}
As a consequence, several approximate $GW$ schemes have been devised.
The most popular approaches are perturbative $GW$, also known as {\GOWO} \cite{Hybertsen_1985a, vanSetten_2013, Bruneval_2012, Bruneval_2013, vanSetten_2015, vanSetten_2018}, eigenvalue self-consistent $GW$ (ev$GW$) \cite{Hybertsen_1986, Shishkin_2007, Blase_2011, Faber_2011} and quasi-particle self-consistent $GW$ (qs$GW$). \cite{Faleev_2004, vanSchilfgaarde_2006, Kotani_2007, Ke_2011, Kaplan_2016}
Within {\GOWO}, the $GW$ self-energy is treated as a perturbation with
respect to a zeroth-order Hamiltonian with a simpler self-energy, such as Hartree-Fock (HF), or a different Hamiltonian altogether, such as a Kohn-Sham Hamiltonian.
The main drawback of {\GOWO} is its dependence on the choice of the starting point, i.e., the zeroth-order Hamiltonian. \cite{Rostgaard_2010,Blase_2011,Ke_2011,Rangel_2016,Kaplan_2016,Caruso_2016} 
Within ev$GW$ the dependence on the starting point is reduced by updating the eigenvalues in a self-consistent field procedure.
However, the orbitals remain those of the zeroth-order Hamiltonian.
Finally, within qs$GW$, the $GW$ self-energy is approximated in such a way that it is both Hermitian and frequency independent.
This allows for a simple self-consistent procedure for both eigenvalues and orbitals eliminating the influence of the starting point.

Although it is known that $GW$ has some shortcomings, they have, until recently, mainly appeared in the strongly correlated regime.~\cite{Romaniello_2009,Romaniello_2012,Berger_2014a,Stan_2015,DiSabatino_2015,DiSabatino_2016,Tarantino_2017,Tarantino_2018,DiSabatino_2020}
However, in two recent articles, \cite{Loos_2018,Veril_2018} we uncovered an important shortcoming of the {\GOWO}, ev$GW$ and qs$GW$ approaches that appears in the weakly correlated regime.
All three approaches suffer from unphysical irregularities and even discontinuities (ev$GW$ and qs$GW$) in important physical quantities such as quasi-particle (QP) energies, neutral excitation energies, and correlation energies.
We showed that the problem could be traced back to the existence of multiple close-lying solutions when the QP energy is close to a pole of the self-energy. \cite{vanSetten_2015,Loos_2018,Veril_2018,Golze_2020}
When the solution switches from one branch to another one it yields an irregularity or discontinuity in the physical observable.
The problem is more severe in ev$GW$ and qs$GW$ because, due to the self-consistency procedure, an irregularity in one QP energy is transferred to all QP energies through the self-consistent procedure.

This problem was again observed in the potential energy surfaces (PES) of diatomic molecules. \cite{Loos_2020}
Accurate results were obtained for the ground-state total energies from the adiabatic-connection fluctuation-dissipation theorem (ACFDT)\cite{Langreth_1975,Gunnarsson_1976,Furche_2005,Toulouse_2009,Toulouse_2010,Angyan_2011,Olsen_2014,Maggio_2016,Holzer_2018,Li_2019,Li_2020} applied to the Bethe-Salpeter equation (BSE) formalism.  \cite{Salpeter_1951,Strinati_1988,Blase_2018,Blase_2020}
However, since the BSE calculations were performed on top of a {\GOWO} calculation, irregularities appeared in the energy curves due to the problem discussed above. As can be anticipated from our discussion above, switching to ev$GW$ or qs$GW$ will not solve the problem. 
Below we will also explicitly show that discontinuities indeed appear in the PES when ev$GW$ or qs$GW$ orbitals and energies are used to calculate the total energy.  
In view of the above, it is desirable to find an alternative approach to {\GOWO}, ev$GW$ and qs$GW$ that does not suffer from this drawback and yields accurate total energies at an affordable computational cost.

In this work we will consider the Coulomb-hole plus screened-exchange (COHSEX) self-energy, which was proposed a long time ago by Hedin, \cite{Hedin_1965,Hybertsen_1986,Hedin_1999} both perturbatively, namely on top of HF, and self-consistently (scCOHSEX).~\cite{Bruneval_2006}
Although the physics inside the COHSEX self-energy is very similar to that included in the $GW$ self-energy, unlike the $GW$ self-energy, it is Hermitian and frequency independent. As a consequence, COHSEX calculations can be done self-consistently using standard numerical techniques (i.e., by simple diagonalization of a Fock-like operator).
A self-consistent COHSEX calculation can also be used as starting point for a {\GOWO} or ev$GW$ calculation.~\cite{Bruneval_2006,Gatti_2007,Vidal_2010,Rangel_2012,Tanwar_2013,Boulanger_2014,Knight_2016}
Such an approach generally yields accurate energy gaps but this would of course suffer from the same irregularities and discontinuities mentioned above.
Thanks to its numerical efficiency COHSEX can be used to perform calculations on large systems.~\cite{Li_2016,Fujita_2018}
The COHSEX approach can also be used to calculate ionic gradients.~\cite{Faber_2015}
Instead, the irregularities and discontinuities in {\GOWO}, ev$GW$ and qs$GW$ could prevent a straightforward calculation of these quantities.
Finally, we note that improvements of the COHSEX method have been proposed.~\cite{Kang_2010}

The main goal of this work is twofold.
We want to show that: (i) physical observables, and in particular PES, obtained within the COHSEX approach do not suffer from irregularities and discontinuities, and (ii) the PES and equilibrium geometries obtained from the BSE using perturbative COHSEX quasi-particle energies (i.e., BSE@COHSEX@HF) are comparable in accuracy to those obtained within BSE@{\GOWO}. We illustrate both points by calculating the PES and equilibrium distances ($R_\text{eq}$) of several diatomic molecules.
Furthermore we want to demonstrate that: (iii) although the COHSEX and {\GOWO} energy gaps are quite different, the influence of this difference on the PES and equilibrium distances is small, and (iv) for the diatomic molecules studied here, perturbative COHSEX, i.e., BSE@COHSEX@HF, yields PES that are in better agreement with the reference values than self-consistent COHSEX, i.e., BSE@scCOHSEX.
Instead, the values of $R_{eq}$ obtained within BSE@scCOHSEX are slightly improved with respect to BSE@COHSEX@HF when compared to the reference data.

The paper is organized as follows. In section \ref{Theory} we describe the theory behind the COHSEX approach and we also briefly discuss the theory of {\GOWO} and partially self-consistent GW methods. We report and discuss our results in section \ref{Results}.
Finally, in section \ref{Conclusions} we draw the conclusions from our work. 

\section{Theory}
\label{Theory}
%
The key variable within many-body perturbation theory is the one-body Green's function $G$.
In the absence of time-dependent fields and at zero temperature, it is defined as
\begin{equation}
\begin{split}
G(\br,\br',\tau) = & -i \Theta(\tau)\mel*{\Psi_0^N}{\hat\psi(\br) e^{-i(\hat{H}^N-E_0^N)\tau} \hat\psi^{\dagger}(\br')}{ \Psi_0^N} 
\\ & + 
i \Theta(-\tau)\mel*{\Psi_0^N}{\hat\psi^{\dagger}(\br')e^{i(\hat{H}^N-E_0^N)\tau}\hat\psi(\br)}{\Psi_0^N},
\label{Eqn:G}
\end{split}
\end{equation}
where $\hat{H}^N$ is the Hamiltonian of the $N$-electron system, $\Psi_0^N$ is its ground-state wave function, $E_0^N$ is the ground-state energy, $\Theta$ is the Heaviside step function, while $\hat\psi^{\dagger}$ and $\hat\psi$ are creation and annihilation operators, respectively.
In practice the one-body Green's function can be obtained from the solution of the following Dyson equation,
\begin{equation}
G(\br,\br',\tau) = G_{\HF}(\br,\br',\tau) + \iint d\br_1 d\br_2 \iint d\tau_1 d\tau_2 G_{\HF}(\br,\br_1,\tau-\tau_1) \Sigma_c(\br_1,\br_2, \tau_2) G(\br_2,\br',\tau_1-\tau_2),
\label{Eqn:Dyson}
\end{equation}
where $G_{\HF}$ is the one-body Green's function within the HF approximation and $\Sigma_c$ is the correlation part of the self-energy which has to be approximated in practical calculations.
\subsection{The COHSEX self-energy}
In this section we discuss the COHSEX self-energy, and, in particular, its correlation part.
We will compare it to the $GW$ self-energy since the two self-energies are similar.
The correlation part of the $GW$ and COHSEX self-energies are given by
\begin{subequations}
\begin{align}
\Sigma_{c}^{GW}(\br,\br',\tau) &= iG(\br,\br',\tau) W_p(\br,\br',\tau+\eta),
\label{Eqn:Sigma_GW}
\\
\Sigma_{c}^{\COHSEX}(\br,\br',\tau) &= iG(\br,\br',\tau) W_p(\br,\br',\omega=0) [\delta(\tau+\eta) + \delta(\tau-\eta)]/2,
\label{Eqn:Sigma_COHSEX}
\end{align}
\end{subequations}
where $W_p = W - v$, is the difference between the screened Coulomb interaction $W$ and the bare Coulomb interaction $v$, $\delta$ is the Dirac delta function, and $\eta$ is a positive infinitesimal that ensures the correct time ordering.
The main difference between the two approximations is that the $GW$ self-energy contains a dynamical (i.e., frequency dependent) $W_p$ while the COHSEX self-energy has a static (i.e., frequency independent) $W_p$.
A Fourier transformation of Eqs.~\eqref{Eqn:Sigma_GW} and \eqref{Eqn:Sigma_COHSEX} yields the following two expressions
\begin{subequations}
\begin{align}
\Sigma_{c}^{GW}(\br,\br',\omega) &= \frac{i}{2\pi} \int d\omega' e^{i\eta\omega'} G(\br,\br',\omega+\omega') W_p(\br,\br',\omega'),
\label{Eqn:GW_FT}
\\
\begin{split}
\Sigma_{c}^{\COHSEX}(\br,\br') &= \frac{i}{2} \qty[ G(\br,\br',-\eta) + G(\br,\br',\eta) ] W_p(\br,\br',\omega=0)
\\ &=
\frac12 \mel*{\Psi_0^N}{\hat\psi(\br) \hat\psi^{\dagger}(\br') - \hat\psi^{\dagger}(\br') \hat\psi(\br)}{\Psi_0^N} W_p(\br,\br',\omega=0),
\label{Eqn:COHSEX_FT}
\end{split}
\end{align}
\end{subequations}
and clearly shows that the COHSEX self-energy is static.
We note that to better understand the screened exchange (SEX) and the Coulomb hole (COH) parts of the COHSEX self-energy it is useful to rewrite Eq.~\eqref{Eqn:COHSEX_FT} according to
\begin{equation}
\begin{split}
\Sigma_{c}^{\COHSEX}(\br,\br') = &
- \mel*{\Psi_0^N}{\hat\psi^{\dagger}(\br') \hat\psi(\br)}{\Psi_0^N} W_p(\br,\br',\omega=0)
\\ &+
\frac12\delta (\br-\br') W_p(\br,\br,\omega=0),
\label{Eqn:Sigma_COHSEX2}
\end{split}
\end{equation}
where we used the anti-commutator relation for the field operators, i.e., $\hat\psi(\br')\hat\psi^{\dagger}(\br)  + \hat\psi^{\dagger}(\br') \hat\psi(\br)= \delta(\br-\br')$.
The first term on the right-hand side of Eq.~\eqref{Eqn:Sigma_COHSEX2} when combined with the HF exchange part of the self-energy, i.e.,
\begin{equation}
\Sigma^{\HF}_x (\br,\br') =  - \langle\Psi_0^N | \hat\psi^{\dagger}(\br') \hat\psi(\br) | \Psi_0^N\rangle v(\br,\br'),
\end{equation}
yields the screened-exchange self-energy.
%
%
The second term on the right-hand side of Eq.~\eqref{Eqn:Sigma_COHSEX2} is the (static) Coulomb-hole self-energy since $W_p(\br,\br,\omega=0)$
is the Coulomb potential at $\br$ due to the Coulomb hole created by an electron present at $\br$.

We can express $W_p$ as
\begin{equation}
W_p(\br,\br',\omega) = \iint d\br_1 d\br_2 v(\br,\br_1) \chi(\br_1,\br_2,\omega) v(\br_2,\br'),
\label{Eqn:Wp}
\end{equation}
where the (reducible) polarizability $\chi$ can be written as
\begin{equation}
\chi(\br,\br',\omega) = \sum_m \left[ \frac{\rho_m(\br)\rho_m(\br')}{\omega - \Omega_m + i\eta} - \frac{\rho_m(\br)\rho_m(\br')}{\omega + \Omega_m - i\eta}\right],
\label{Eqn:chi}
\end{equation} 
in which $\Omega_m$ is a neutral excitation energy and $\rho_m$ the corresponding transition density.
The latter is defined as
\begin{equation}
\rho_m(\br) = \sum_{i}^{\text{occ}}\sum_{a}^{\text{virt}} (\bX+\bY)_{ia}^{m} \phi_i (\br) \phi_a (\br)
\end{equation}
where $\phi_p$ are either the (real-valued) HF spatial orbitals $\phi^{\HF}_p$ (for a COHSEX@HF calculation) or the (real-valued) scCOHSEX spatial orbitals $\phi^{\COHSEX}_p$, i.e., the eigenfunctions of the COHSEX Hamiltonian $\hat{H}^{\COHSEX} = \hat{H}^{\HF} + \hat{\Sigma}_c^{\COHSEX}$.
In the following, the index $m$ labels the single excitations; $i$ and $j$ are occupied orbitals; $a$ and $b$ are unoccupied orbitals, while $p$, $q$, $r$, and $s$ indicate arbitrary orbitals.

The neutral excitation energies $\Om{m}$ and the transition amplitudes $(\bX+\bY)_{m}^{ia}$ are obtained from a random-phase approximation (RPA) calculation:
\begin{equation}
\label{eq:LR}
	\begin{pmatrix}
		\mathrm{\bf{A}}	&	\bB	\\
		-\bB	&	-\bA	\\
	\end{pmatrix}
	\begin{pmatrix}
		\bX_m	\\
		\bY_m	\\
	\end{pmatrix}
	=
	\bOm_m
	\begin{pmatrix}
		\bX_m	\\
		\bY_m	\\
	\end{pmatrix},
\end{equation}
where $(\bX_m,\bY_m)^T$ is the eigenvector that corresponds to $\bOm_m$, and
\begin{subequations}
\begin{align}
\label{eq:RPA_A}
\text{A}_{ia,jb} & = \delta_{ij} \delta_{ab} (\epsilon_a - \epsilon_i) + 2 (ia|jb),
\\
\label{eq:RPA_B}
\text{B}_{ia,jb} & = 2 (ia|bj),
\end{align}
\end{subequations}
where $\epsilon_p$ are either the HF orbital energies $\epsilon^{\HF}_p$ (for a COHSEX@HF calculation) or the scCOHSEX orbital energies $\epsilon^{\COHSEX}_p$ (i.e., the eigenvalues of $\hat{H}^{\COHSEX}$), and $(pq|rs)$ are the bare two-electron integrals defined as
\begin{equation}
(pq|rs) = \iint d\br d\br' \phi_p(\br) \phi_q(\br) v(\br,\br') \phi_r(\br') \phi_s(\br').
\end{equation}

While the $GW$ self-energy is non-Hermitian and frequency dependent,
the COHSEX self-energy is both static and Hermitian as can be verified from the expression one obtains by inserting Eq.~\eqref{Eqn:chi} into Eq.~\eqref{Eqn:COHSEX_FT} (with $W_p$ given by (\ref{Eqn:Wp})):
\begin{multline}
\Sigma_{c}^{\COHSEX}(\br,\br') = \qty[\mel*{\Psi_0^N}{\hat\psi^{\dagger}(\br')\hat\psi(\br)}{ \Psi_0^N} 
-\mel*{\Psi_0^N}{\hat\psi(\br) \hat\psi^{\dagger}(\br')}{\Psi_0^N}  ] 
\\ \times
\iint d\br_1 d\br_2 v(\br,\br_1) \sum_m \frac{\rho_m(\br_1)\rho_m(\br_2)}{\Omega_m} v(\br_2,\br').
\end{multline}
Moreover, it is important to note that the COHSEX self-energy has no poles.
More precisely, its denominator never vanishes since the $\Omega_m$ are real and positive for finite systems.
Owing to the Hermiticity and frequency-independence of the COHSEX self-energy, $\Psi_0^N$ can be represented by a single Slater determinant.
Following the Slater-Condon rules the matrix elements in the above equation can then be rewritten as sums of products of orbitals.
We obtain
\begin{multline}
\Sigma_{c}^{\COHSEX}(\br,\br') = 2\left[\sum_i^{\text{occ}} \phi_i(\br) \phi_i(\br') - \sum_a^{\text{virt}} \phi_a(\br) \phi_a(\br')\right]
\\ \times
\iint d\br_1 d\br_2 v(\br,\br_1) \sum_m \frac{\rho_m(\br_1)\rho_m(\br_2)}{\Om{m}} v(\br_2,\br').
\end{multline}
%

The matrix element $\Sigma_{c,pq}^{\COHSEX} = \mel*{\phi_p}{\Sigma_c^{\COHSEX}}{\phi_q}$ can now be written as
\begin{equation}
\Sigma^{\COHSEX}_{c,pq} = 2 \sum_m \qty[ \sum_{i}^\text{occ} \frac{[pi|m][qi|m]}{ \Om{m}} - \sum_{a}^\text{virt} \frac{[pa|m][qa|m]}{\Om{m}} ],
\end{equation}
where the screened two-electron integrals are defined as
\begin{equation}
[pq|m] = \sum_{ia} (pq|ia) (\bX+\bY)_{m}^{ia}.
\end{equation}

%
When COHSEX is performed using first-order perturbation with respect to HF, the perturbation is given by $\hat{H}^{\COHSEX} - \hat{H}^{\HF} = \hat{\Sigma}^{\COHSEX}_c$.
The perturbative COHSEX orbital energies can thus be obtained from
\begin{equation}
\epsilon_p^{\COHSEX} = \epsilon^{\HF}_p + \Sigma^{\COHSEX}_{c,pp}.
\end{equation}
Instead, within scCOHSEX both the eigenvalues and eigenfunctions of the COHSEX Hamilonian have to be calculated repeatedly
until a self-consistent result is obtained.

\subsection{{\GOWO}}
Given the difficulty of evaluating the $GW$ self-energy mentioned before one often uses a perturbative approach called {\GOWO} in which the 
self-energy is calculated perturbatively with respect to a simpler zeroth-order Hamiltonian, such as a self-energy for which a self-consistent solution is more easily obtained.
In this work we will use the HF Green's function as our zeroth-order Green's function.
Its spectral representation is given by
\begin{equation}
G_{\HF}(\br,\br',\omega) = \sum_p \frac{\phi^{\HF}_p(\br)\phi^{\HF}_p(\br')}{\omega - \epsilon^{\HF}_p - i\eta\sign(\mu - \epsilon^{\HF}_p)},
\end{equation}
with $\mu$ the chemical potential. 
Within the {\GOWO} approximation, the frequency integral in Eq.~\eqref{Eqn:GW_FT} can be performed analytically and one obtains the following matrix elements of the  {\GOWO} self-energy,
\begin{equation} \label{Eqn:SigGW}
\Sigma^{G_0W_0}_{c,pq}(\omega) = 2 \sum_m \qty[ \sum_{i}^\text{occ} \frac{[pi|m]^{\HF}[qi|m]^{\HF}}{\omega - \epsilon^{\HF}_i + \Om{m}^{\HF} - i \eta} + \sum_{a}^\text{virt} \frac{[pa|m]^{\HF}[qa|m]^{\HF}}{\omega - \epsilon^{\HF}_a - \Om{m}^{\HF} + i \eta}],
\end{equation}
where the superscript in $\Omega_m^{\HF}$ and $[pq|m]^{\HF}$ indicates that these quantities are obtained from HF eigenvalues and orbitals.
Contrary to the COHSEX self-energy, the above self-energy is dynamical and has poles.
The QP energies can then be obtained from the poles of $G$ obtained by solving the Dyson equation \eqref{Eqn:Dyson} (in frequency space) with the above self-energy.
This yields the so-called QP equation,
\begin{equation}
\omega = \epsilon^{\HF}_p + \Re[\Sigma^{G_0W_0}_{c,pp}(\omega)].
\end{equation}
Due to the frequency dependence of the self-energy, the {\GOWO} QP equation has, in general, multiple solutions $\epsilon_{p,s}^{G_0W_0}$.
The solution $\epsilon_{p}^{G_0W_0} \equiv \epsilon_{p,s=0}^{G_0W_0}$ with the largest spectral weight $Z_p(\epsilon_{p,s=0}^{G_0W_0})$ with
\begin{equation} \label{Eqn:Z}
Z_{p}(\omega) = \qty[ 1 - \frac{\Re[\Sigma^{G_0 W_0}_{c,pp}(\omega)]}{\partial\omega} ]^{-1},
\end{equation}
is called the QP solution (or simply quasi-particle), while the other solutions ($s > 0$) are called satellites and share the rest of the spectral weight.
In practice the QP equation is often simplified by Taylor expanding the self-energy to first order around $\epsilon^{\HF}_p$.
The result is the so-called linearized QP equation given by
\begin{equation}
\epsilon_p^{G_0W_0} = \epsilon^{\HF}_p + Z_p(\epsilon^{\HF}_p)\Re[\Sigma^{G_0W_0}_{c,pp}(\epsilon^{\HF}_p)].
\end{equation}
When the self-energy has poles close to a solution of the QP equation the above linearization is not justified.
Moreover, it leads to irregularities in physical observables such as PES. This can be understood as follows.

Although the self-energy in the linearized QP equation is independent of the frequency its denominator could still vanish.
This happens when $\epsilon^{\HF}_p = \epsilon^{\HF}_i - \Omega^{\HF}_m$ or when $\epsilon^{\HF}_p = \epsilon^{\HF}_a + \Omega^{\HF}_m$.
When calculating a single QP for a single configuration of an atom or a molecule it is not very probable that such an event occurs.
However, when a large number of QPs and/or configurations is considered, e.g., when calculating a PES, it becomes inevitable.
As an example, let us consider the simplest PES, namely the variation of the total energy of a diatomic molecule as a function of the interatomic distance $R$.
In such a case, $\epsilon^{\HF}_p$ and $\Omega^{\HF}_m$ could be considered functions of $R$ and the conditions that the self-energy has a vanishing denominator can be written as
\begin{subequations}
\begin{align}
\epsilon^{\HF}_p (R) &= \epsilon^{\HF}_i (R) - \Omega^{\HF}_m (R),
\label{Eqn:cond1}
\\
\epsilon^{\HF}_p (R) &= \epsilon^{\HF}_a (R) + \Omega^{\HF}_m (R).
\label{Eqn:cond2}
\end{align}
\end{subequations}
Therefore, $\Sigma^{G_0W_0}_{c,pp}[\epsilon^{\HF}_p(R)]$ can be considered an implicit function of $R$ that has poles.
From the above conditions it is clear that in a region equal to $2\Omega^{\HF}_0 (R)  +  \epsilon^{\HF}_{\text{LUMO}} (R) -  \epsilon^{\HF}_{\text{HOMO}} (R)$ around the Fermi level no poles can occur, where $\Omega^{\HF}_0$ is the smallest neutral excitation energy and $\epsilon^{\HF}_{\text{LUMO}}$ and $\epsilon^{\HF}_{\text{HOMO}}$ are the HF energies of the lowest unoccupied molecular orbital (LUMO) and the highest occupied molecular orbital (HOMO), respectively.
However, since, in general, the variation with respect to $R$ of the left- and right-hand sides of Eqs.~\eqref{Eqn:cond1} and \eqref{Eqn:cond2} is different, it is unavoidable that outside of this range one of the two above conditions is met for some values $R=R_p$.  In the vicinity of these $R_p$ values the self-energy [see Eq.~\eqref{Eqn:SigGW}] and its corresponding renormalization factor $Z_p$ [see Eq.~\eqref{Eqn:Z}] vary rapidly leading to irregularities in the QP energies and, hence, in the PES.
We note that satisfaction of either Eq.~\eqref{Eqn:cond1} or Eq.~\eqref{Eqn:cond2} will ensure an irregularity and that they can never be met simultaneously.
\subsection{Partially self-consistent $GW$}
The main drawback of the {\GOWO} approach is its dependence on the starting point, i.e,. the orbitals and energies of the zeroth-order Hamiltonian.
Since, as mentioned before, from a numerical point of view, fully self-consistent $GW$ is nontrivial, so-called partial self-consistent $GW$ methods have been developed to reduce or eliminate the starting-point dependence.
Within ev$GW$ one only updates the eigenvalues in the self-energy while in qs$GW$ one symmetrizes the {\GOWO} self-energy according to
\begin{equation}
\Sigma^{\qsGW}_{c,pq} = \frac12\Re\qty[\Sigma^{G_0W_0}_{c,pq}(\epsilon^{\qsGW}_p) 
+ \Sigma^{G_0W_0}_{c,pq}(\epsilon^{\qsGW}_q) ].
\end{equation}
%
The above self-energy is frequency-independent and Hermitian and is, hence, suitable for a standard self-consistent procedure.
Therefore, in this partially self-consistent scheme both the eigenvalues and orbitals are updated.

However, the ev$GW$ and qs$GW$ approaches suffer from the same problem as {\GOWO} since the self-energies have poles when considered as (implicit) functions of the geometry.
In fact the problem is even more severe since, due to the self-consistent procedure, an irregularity in one QP energy is transferred to all the other QP energies.
As a consequence, in some regions of the geometry space, there is more than one branch of solutions and discontinuities appear when a solution switches from one branch to another.

To be more precise let us consider a diatomic molecule.
For certain internuclear distances $R_{d}$ two solutions of the QP equation can have equal weight but not equal energies.
Just before this point $R_{d}$ one of the two solutions will be picked up in the self-consistent procedure, since it has a slightly larger weight than the other solution.
Instead, just after $R_{d}$ the roles are reversed and the other solution will be picked up because it is now the one with the slightly larger weight.
Since the two solutions have different energies there is a sudden jump of the QP energy at $R_{d}$ causing a discontinuity.
This scenario occurs whenever a solution of the QP equation lies close to a pole of the self-energy.~[\citenum{Loos_2018,Veril_2018}]
It can happen for any state, occupied or virtual, except those close to the Fermi level since it can be shown explicitly that the self-energy has no poles in an energy region with a width that is equal to the HOMO-LUMO energy gap plus twice the lowest neutral excitation energy.
More details and analysis of the origin of irregularities and discontinuities in $GW$ approaches can be found in Refs.~[\citenum{Loos_2018,Veril_2018}].

%
\subsection{Correlation energy}
We calculate the correlation energies at the BSE level using an approach based on the ACFDT. \cite{Langreth_1975,Gunnarsson_1976,Furche_2005} 
We note that the ACFDT formalism is formally derived for a local potential, while here the potential, i.e., the self-energy, is non-local.
We strictly follow the ACFDT procedure described in Ref.~\citenum{Loos_2020} and the details can be found there.
For the sake of completeness we briefly discuss some details of the calculation of the BSE total energy.
The main difference with Ref.~\citenum{Loos_2020} is that the QP energies and orbitals appearing in the equations below are those pertaining to the COHSEX self-energy instead of the {\GOWO} self-energy.
Finally, we note that the approach described below is compatible only with a quasiparticle $GW$ calculation but not fully self-consistent $GW$, since it requires orbitals and orbital energies as input.

Within the ACFDT formalism, the BSE correlation energy can be written as an integral over the coupling constant $\lambda$ which adiabatically connects
the noninteracting system ($\lambda=0$) with the fully interacting system ($\lambda=1$) according to \cite{Olsen_2014,Maggio_2016,Holzer_2018,Li_2019,Li_2020,Loos_2020}
\begin{equation}
\label{Eqn:Ec_BSE}
E_c^{\text{BSE}} = \frac12 \int_0^1 \text{Tr} \left(\mathbf{K}\mathbf{P}^{\lambda}\right) d\lambda
\end{equation}
where the polarizability matrix $\mathbf{P}^{\lambda}$ is given by
\begin{equation}
\mathbf{P}^{\lambda} = 
	\begin{pmatrix}
		\bY^{\lambda}(\bY^{\lambda})^{T}	&	\bY^{\lambda}(\bX^{\lambda})^{T}	\\
		\bX^{\lambda}(\bY^{\lambda})^{T}	&	\bX^{\lambda}(\bX^{\lambda})^{T}	\\
	\end{pmatrix}
	-
	\begin{pmatrix}
		\mathbf{0}	&	\mathbf{0}		\\
		\mathbf{0}		&	\mathbf{1}	\\
	\end{pmatrix}
\end{equation}
with $\bX^{\lambda}$ and $\bY^{\lambda}$ solutions of
\begin{equation}
	\begin{pmatrix}
		\bA^{\lambda,\text{BSE}}	&	\bB^{\lambda,\text{BSE}}	\\
		-\bB^{\lambda,\text{BSE}}	&	-\bA^{\lambda,\text{BSE}}	\\
	\end{pmatrix}
	\begin{pmatrix}
		\bX^{\lambda}_m	\\
		\bY^{\lambda}_m	\\
	\end{pmatrix}
	=
	\bOm_m^{\lambda}
	\begin{pmatrix}
		\bX_m^{\lambda}	\\
		\bY_m^{\lambda}	\\
	\end{pmatrix},
\end{equation}
where
\begin{align}
\label{Eq:BSE}
\text{A}^{\lambda,\text{BSE}}_{ia,jb} & = \delta_{ij} \delta_{ab} (\epsilon_a - \epsilon_i) + \lambda \left[2 (ia|jb) - W^{\lambda}_{ij,ab}\right],
\\
\text{B}^{\lambda,\text{BSE}}_{ia,jb} & = \lambda \left[ 2 (ia|bj) - W^{\lambda}_{ib,aj} \right],
\end{align}
with
\begin{equation}
W^{\lambda}_{pq,rs} = \iint d\br d\br' \phi_p(\br) \phi_q(\br) W^{\lambda}(\br,\br',\omega=0) \phi_r(\br') \phi_s(\br'),
\end{equation}
Finally, the interaction kernel $\mathbf{K}$ is given by
\begin{equation}
\mathbf{K} = 
\begin{pmatrix}
		\tilde\bA^{\text{BSE}}	&	\bB^{\lambda=1,\text{BSE}}	\\
		\bB^{\lambda=1,\text{BSE}}	&	\tilde\bA^{\text{BSE}}	\\
	\end{pmatrix}
\end{equation}
with $\tilde{\text{A}}^{\text{BSE}}_{ia,jb} = 2 (ia|bj)$.
We note that Eq.~\eqref{Eqn:Ec_BSE} is referred to as ``extended Bethe-Salpeter (XBS)'' in Ref.~\citenum{Holzer_2018} .
An important point to make here is that, in contrast to Kohn-Sham density-functional theory where the electron density is fixed along the adiabatic path, \cite{Langreth_1975,Gunnarsson_1976} the density is not maintained in the present BSE formalism as the coupling constant varies. 
Therefore, an additional contribution to Eq.~\eqref{Eqn:Ec_BSE} originating from the variation of the Green's function along the adiabatic connection path should be, in principle, added. \cite{Hesselmann_2011}
However, as it is commonly done, \cite{Toulouse_2009, Toulouse_2010, Colonna_2014, Holzer_2018} we shall neglect it in the present study.

The BSE total energy $E^{\text{BSE}}$ of the system can then be written as
\begin{equation}
E^{\text{BSE}} = E^{\text{nuc}} + E^{\HF} + E_c^{\text{BSE}}
\end{equation}
where $E^{\text{nuc}}$ and $E^{\HF}$ are the nuclear energy and the HF energy, respectively.
We note that for a BSE@scCOHSEX calculation $E^{\HF}$ is calculated with the scCOHSEX orbitals.

\section{Results}
\label{Results}

All systems under investigation have a closed-shell singlet ground state.
Hence, the restricted HF formalism has been systematically employed in the present study.
The infinitesimal $\eta$ is set to zero for all calculations. 
The numerical integration required to compute the correlation energy along the adiabatic path (see Eq.~\eqref{Eqn:Ec_BSE}) is performed with a 21-point Gauss-Legendre quadrature.
All the calculations have been performed with the software \texttt{QuAcK}, \cite{QuAcK} freely available on \texttt{github}.
\texttt{QuAcK} uses Gaussian type orbitals and its implementation closely follows that of MOLGW.~\cite{Bruneval_2016}
In particular, the frequency integral in the {\GOWO} self-energy is done exactly, i.e., we solve Eq.~\eqref{eq:LR} and use the neutral excitation energies in Eq.~\eqref{Eqn:SigGW}.
The threshold for the convergence of the quasi-particle energies was set to 10$^{-6}$ Ha and 10$^{-5}$ Ha in the HF and scCOHSEX calculations, respectively.
We have used the DIIS technique to accelerate convergence. ~\cite{Pulay_1980,Pulay_1982}
As one-electron basis sets, we employ the Dunning family (cc-pVXZ) defined with cartesian Gaussian functions.
Finally, we note that we diagonalize Eq.~\eqref{eq:LR} which is a $(OV)\times (OV)$ matrix (where $O$ and $V$ are the number of occupied and virtual orbitals).
Because a complete diagonalization scales as $N^3$, with $N$ the number of electrons, for a matrix of size $N\times N$, computing the screening in such a way scales as $(OV)^3$, i.e., $N^6$. Several techniques exist to improve the scaling of the calculation of the screening.~\cite{Berger_2010,Berger_2012a,Berger_2012b,Duchemin_2019,Duchemin_2020}

\subsection{Irregularities and discontinuities in {\GOWO}, ev$GW$, and qs$GW$}
\begin{figure*}[h]
	\includegraphics[width=\linewidth]{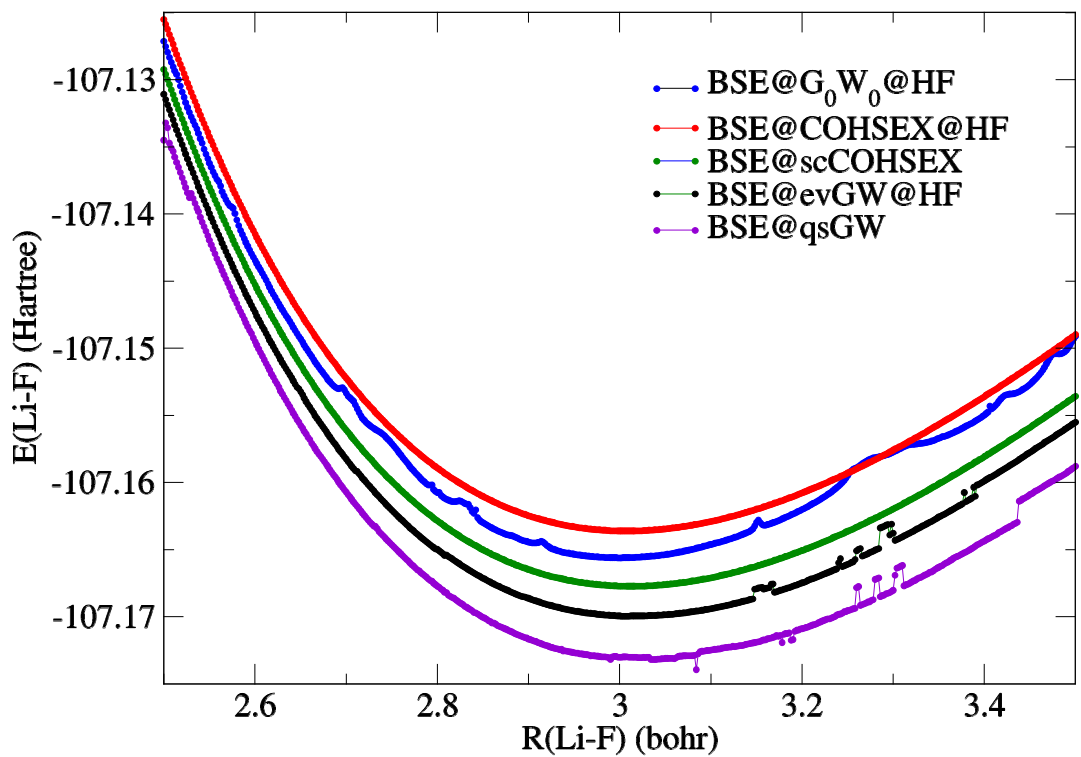}
	\caption{
	\label{Fig:LiF_pvdz} The BSE total energy of the LiF molecule in the cc-pVDZ basis as a function of the internuclear distance. The calculations were done at intervals of 0.002 bohr.
}
\end{figure*}

We have previously described in detail the problem of irregularities and discontinuities in physical observables obtained from {\GOWO} and partially self-consistent $GW$ approaches.~\cite{Loos_2018,Veril_2018}
 Here we want to remind the reader that these problems are also present in total energy calculations and we want to show that, instead, there are no such problems in the COHSEX method.
In Fig.~\ref{Fig:LiF_pvdz} we report the BSE total energy of the LiF molecule as a function of the interatomic distance in the vicinity of its equilibrium distance. 
The BSE correlation energy is calculated on top of {\GOWO}@HF, COHSEX@HF, ev$GW$@HF, qs$GW$, and scCOHSEX.
We used a relatively small basis set, namely Dunning's cc-pVDZ basis, since for larger basis sets the qs$GW$ approach does not yield converged results for many values of $R$. This, however, does not change the conclusions of this section.
We note that within qs$GW$ the entire set of energies and orbitals is updated at each iteration.
We see that all four results are within a range of about $10$ mHartree.
However, the PES obtained from BSE@{\GOWO}@HF shows irregularities while the PES obtained from BSE@ev$GW$@HF and BSE@qs$GW$ show discontinuities. In fact, the different branches of solutions can clearly be seen, especially around 3.4 bohr.
Instead, the BSE total energies obtained on top of a COHSEX calculation, i.e., BSE@COHSEX@HF and BSE@scCOHSEX, yield a PES that is a smooth function of the interatomic distance.

Finally, we note that including self-consistency in COHSEX and $GW$ tends to lower the total energies and that including self-consistency for both QP energies and orbitals lowers the total energy more than just including self-consistency for the QP energies.
Moreover, the effect of self-consistency on the total energies in COHSEX, going from COHSEX@HF to scCOHSEX, is roughly identical to the effect on $GW$, going from {\GOWO}@HF to ev$GW$@HF.
\subsection{Ground-state PES}
\begin{figure*}[h]
	\includegraphics[width=\linewidth]{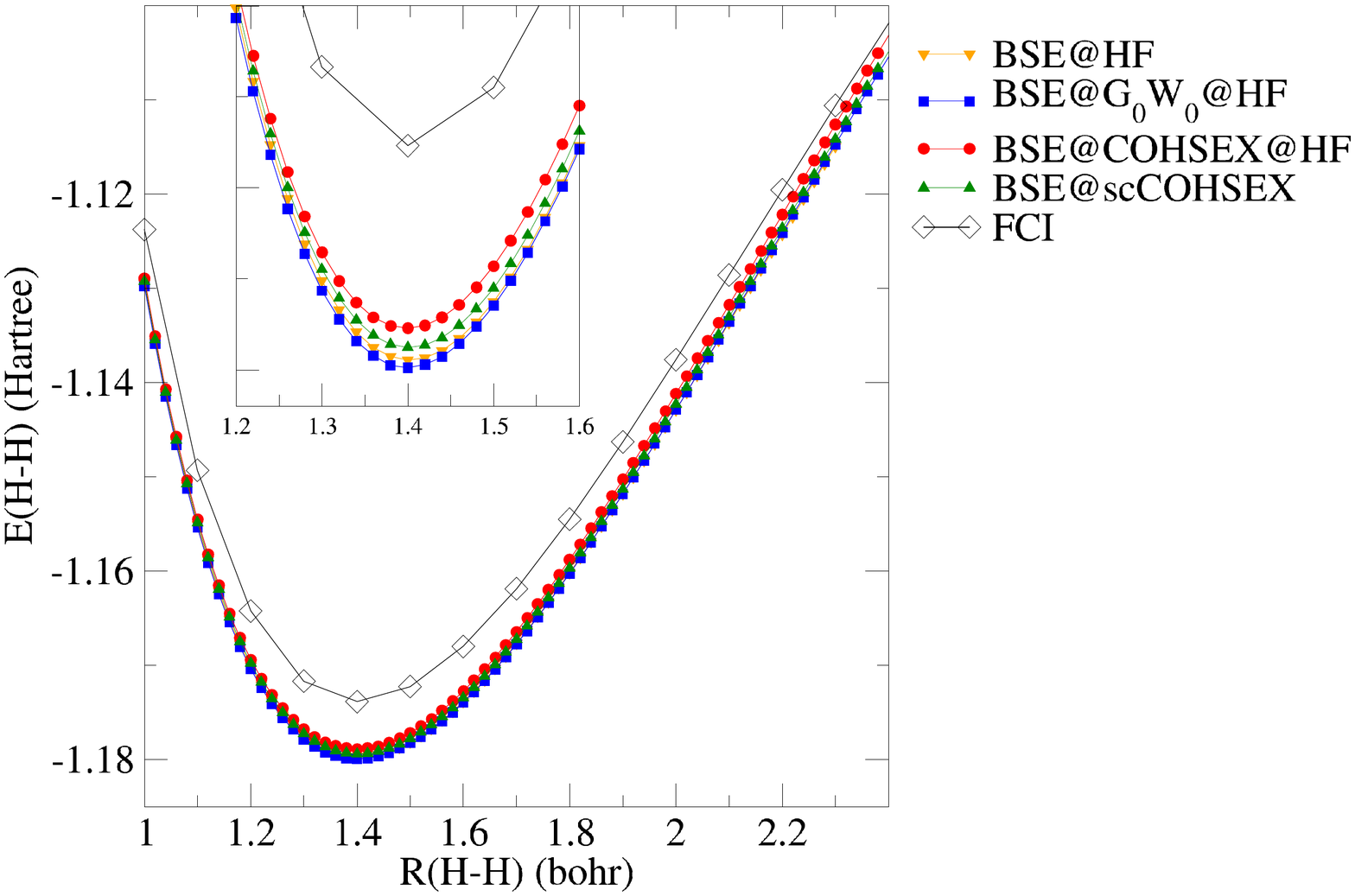}
	\caption{
	\label{Fig:H2_pvqz} The total energy of the H$_2$ molecule in the cc-pVQZ basis as a function of the internuclear distance. 
}
\end{figure*}

\begin{figure*}[h]
	\includegraphics[width=\linewidth]{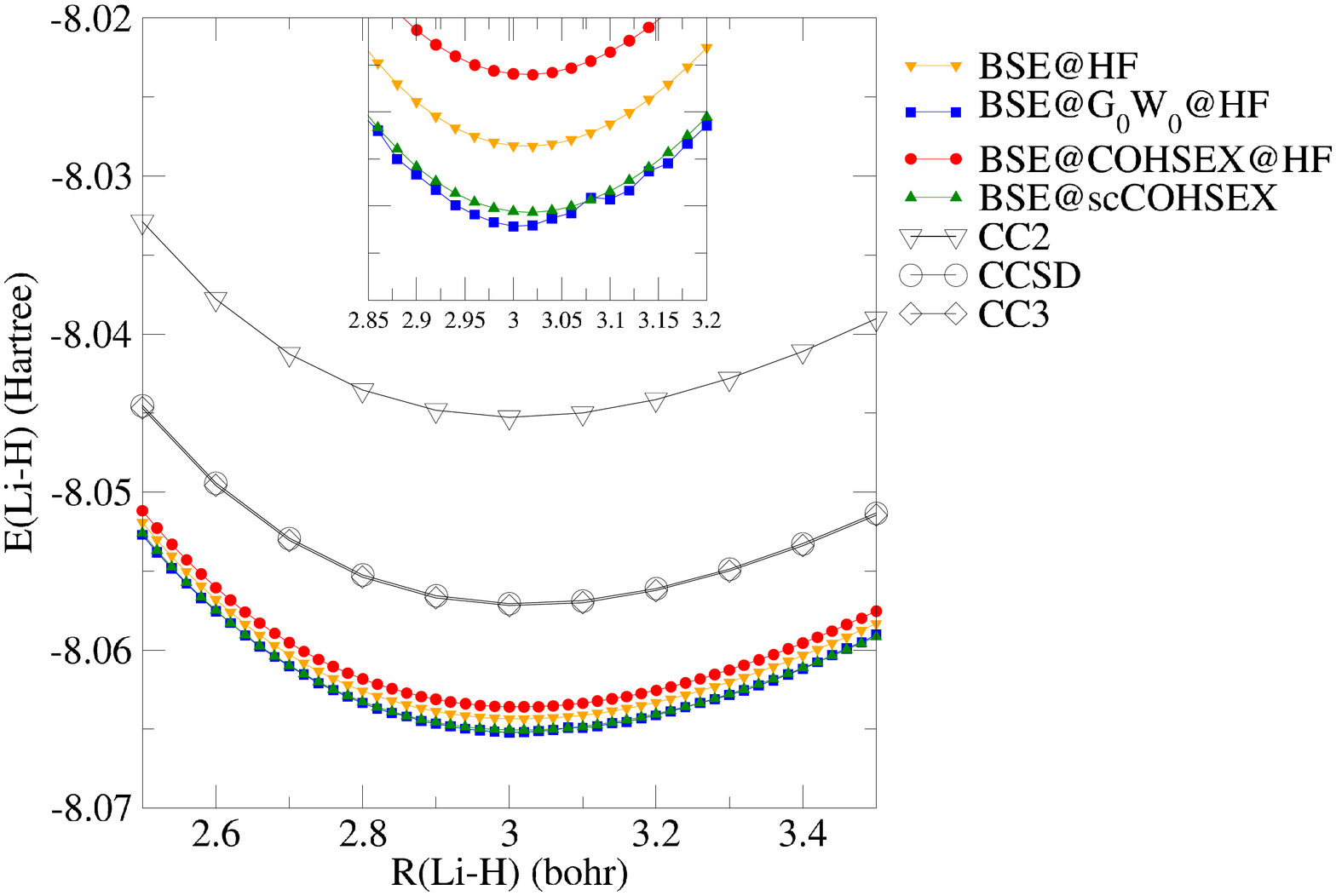}
	\caption{
	\label{Fig:LiH_pvqz} The total energy of the LiH molecule in the cc-pVQZ basis as a function of the internuclear distance. 
}
\end{figure*}

\begin{figure*}[h]
	\includegraphics[width=\linewidth]{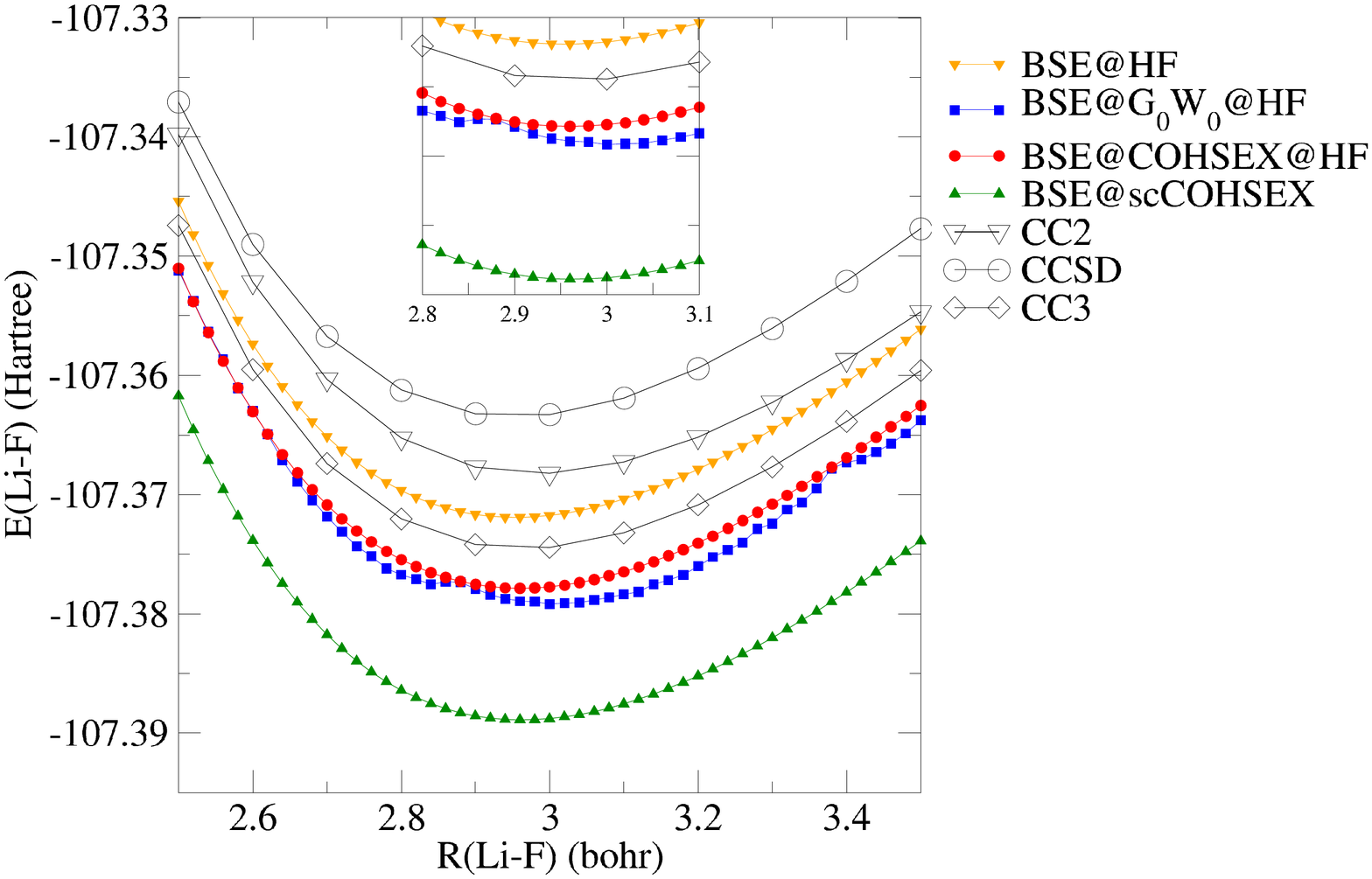}
	\caption{
	\label{Fig:LiF_pvqz} The total energy of the LiF molecule in the cc-pVQZ basis as a function of the internuclear distance. 
}
\end{figure*}

\begin{figure*}[h]
	\includegraphics[width=\linewidth]{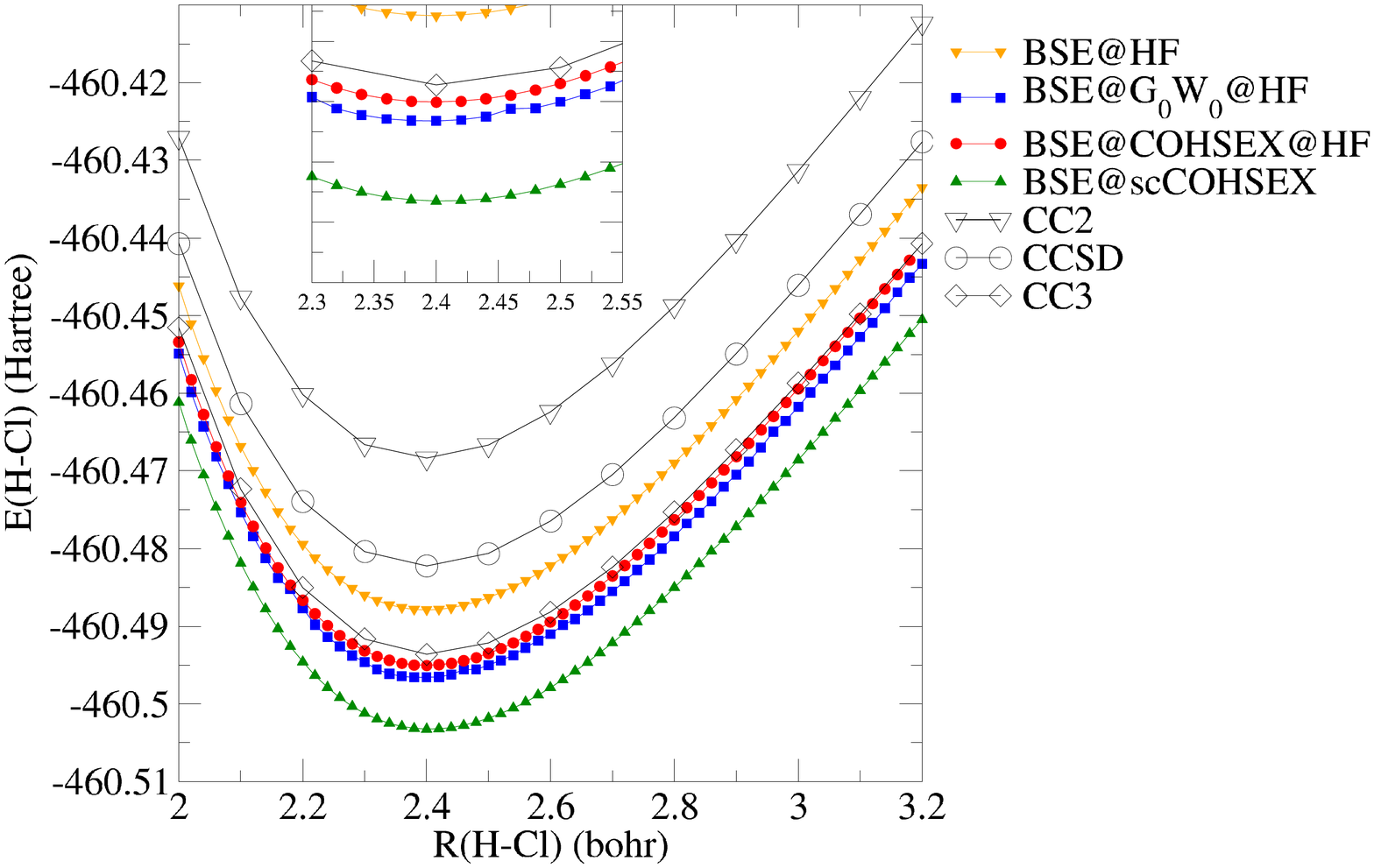}
	\caption{
	\label{Fig:HCl_pvqz} The total energy of the HCl molecule in the cc-pVQZ basis as a function of the internuclear distance. 
}
\end{figure*}

\begin{figure*}[h]
	\includegraphics[width=\linewidth]{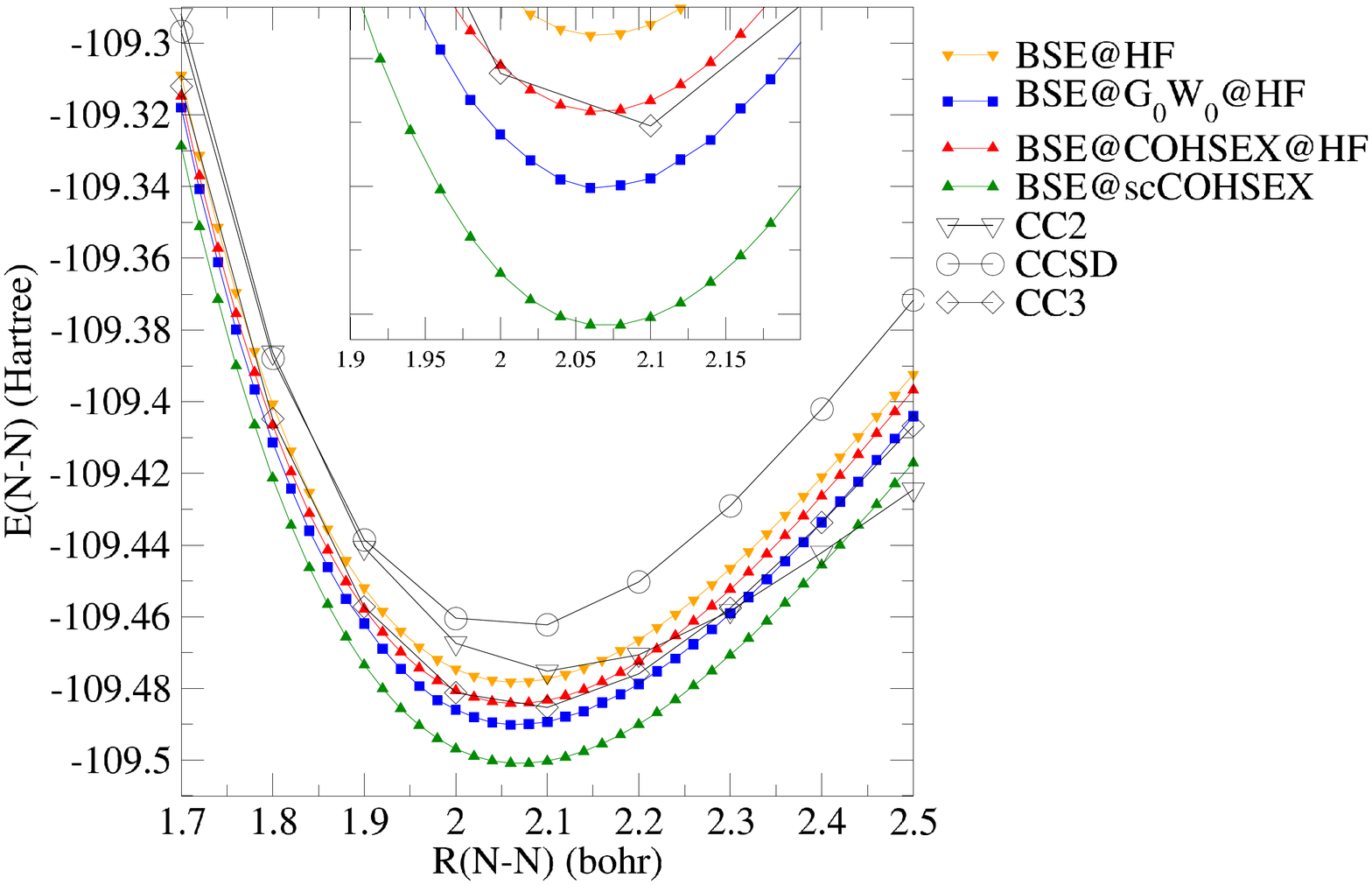}
	\caption{
	\label{Fig:N2_pvqz} The total energy of the N$_2$ molecule in the cc-pVQZ basis as a function of the internuclear distance. 
}
\end{figure*}

\begin{figure*}[h]
	\includegraphics[width=\linewidth]{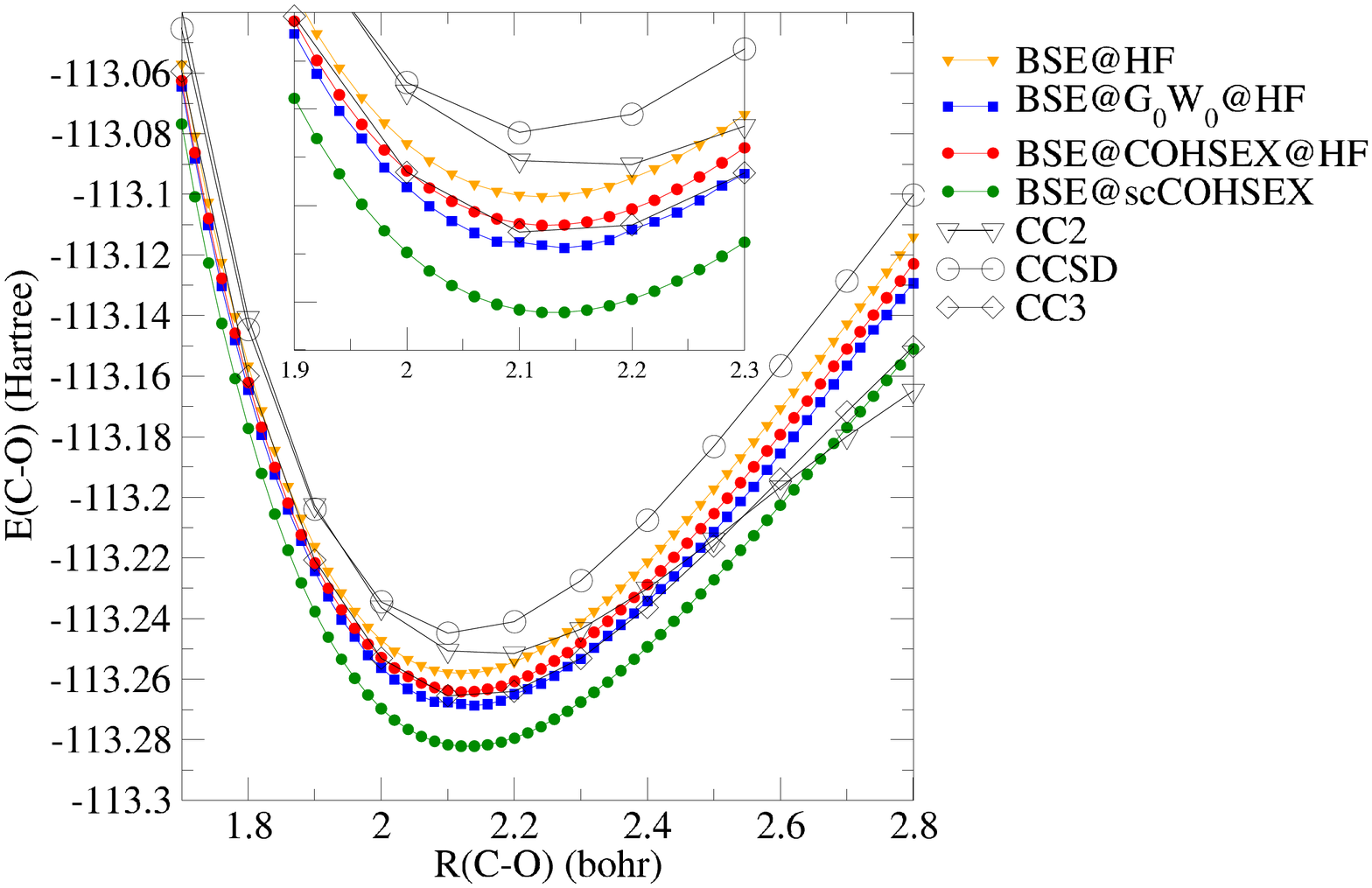}
	\caption{
	\label{Fig:CO_pvqz} The total energy of the CO molecule in the cc-pVQZ basis as a function of the internuclear distance. 
}
\end{figure*}

\begin{figure*}[h]
	\includegraphics[width=\linewidth]{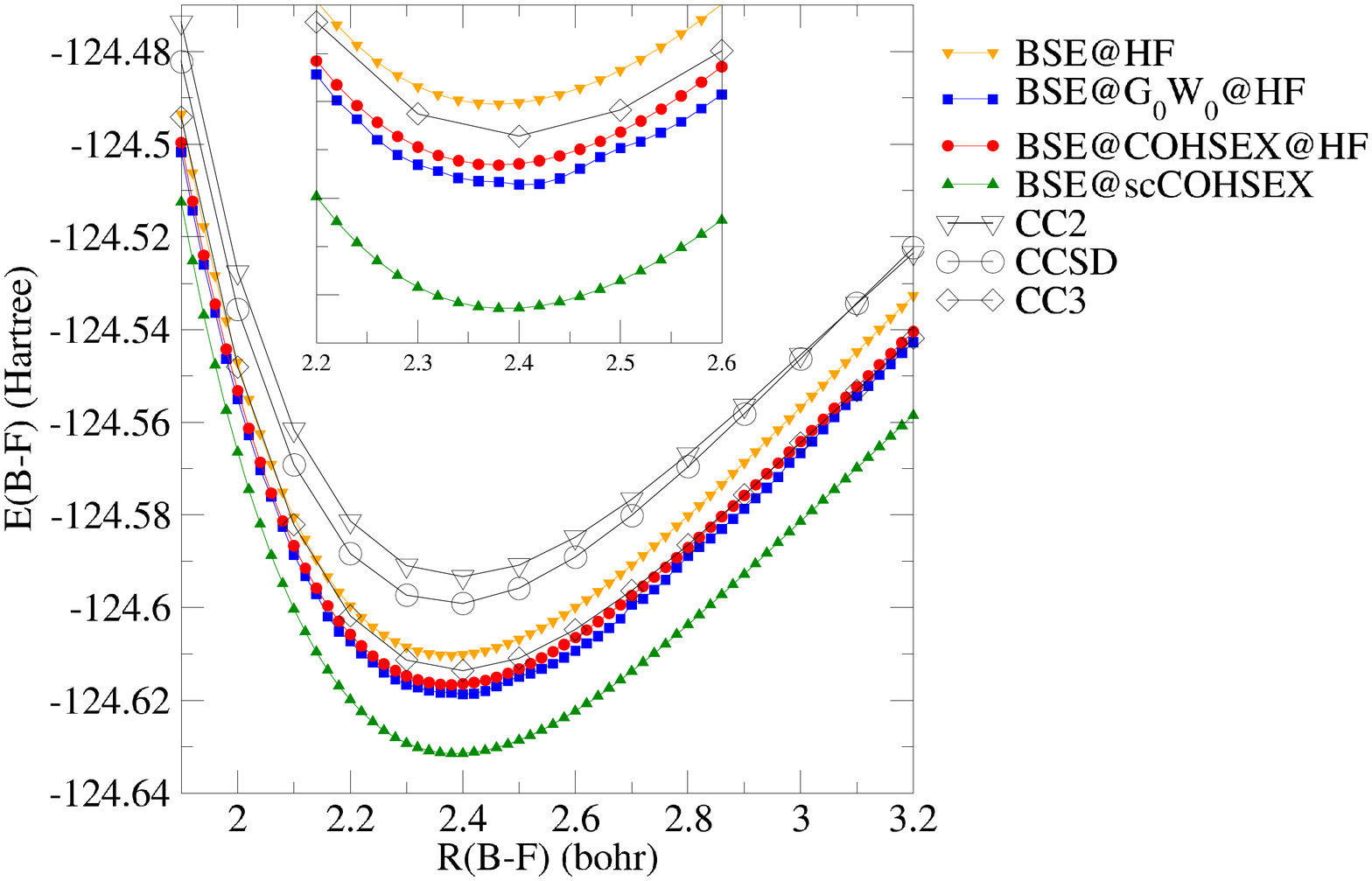}
	\caption{
	\label{Fig:BF_pvqz} The total energy of the BF molecule in the cc-pVQZ basis as a function of the internuclear distance. 
}
\end{figure*}

\begin{figure*}[h]
	\includegraphics[width=\linewidth]{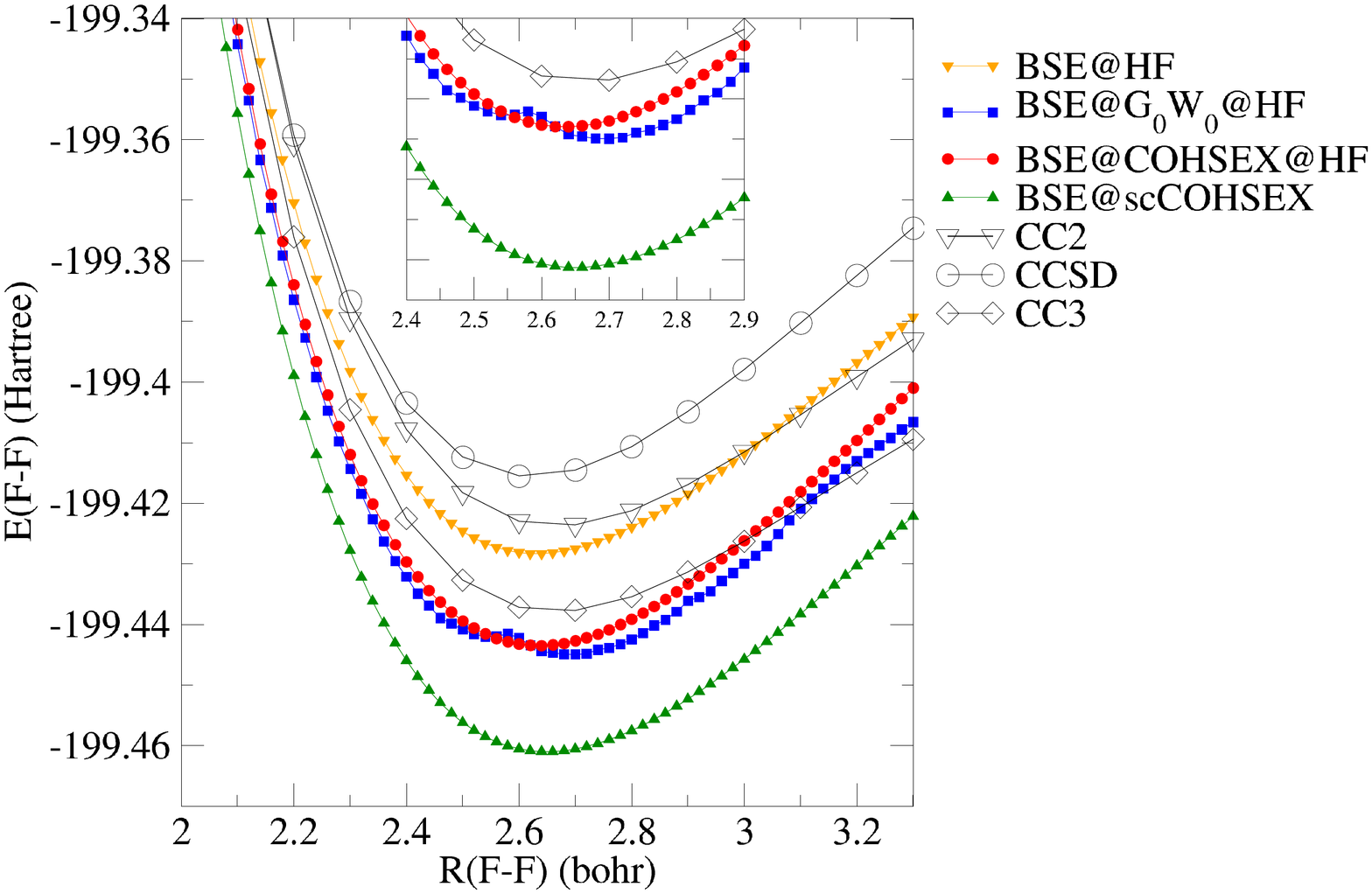}
	\caption{
	\label{Fig:F2_pvqz} The total energy of the F$_2$ molecule in the cc-pVQZ basis as a function of the internuclear distance. 
}
\end{figure*}

In Figs.~\ref{Fig:H2_pvqz}-\ref{Fig:F2_pvqz} we report the BSE total energies on top of COHSEX@HF, {\GOWO}@HF and scCOHSEX as a function of the interatomic distance around the equilibrium distance for the following diatomic molecules:
H$_2$, LiH, LiF, HCl, N$_2$, CO, BF, and F$_2$, respectively.
They are the same molecules that were studied in Ref.~\citenum{Loos_2020}. We also use the same basis set, namely Dunning's cc-pVQZ.
For comparison we also report the PES obtained with the coupled cluster (CC) methods of increasing accuracy: CC2 \cite{Christiansen_1995}, CCSD \cite{Purvis_1982}, CC3. \cite{Christiansen_1995_2}. At the equilibrium distance the CC3 approach has been shown to yield total energies that are very close to those obtained with higher-order CC approaches, such as CCSDT and CCSDT(Q). \cite{Loos_2020}
Therefore, we can consider it to be the reference method.
We also compare to the PES obtained within BSE@HF in which the BSE is solved using HF orbital energies.

In the only case for which we have an exact result (for the given basis set), namely the H$_2$ PES obtained from full configuration interaction (FCI), all BSE total energies are roughly the same. We also note that no irregularities are visible in the BSE@{\GOWO}@HF curve.
In the case of LiH, the second smallest molecule in the set, an irregularity appears in the BSE@{\GOWO}@HF curve around 3.08 bohr.
We also observe that the smooth BSE@COHSEX@HF total-energy curves are closest to the reference CC3 values, while BSE@scCOHSEX and BSE@{\GOWO}@HF yield almost identical energies. 
For the LiF molecule there are large irregularities in the PES obtained within BSE@{\GOWO}@HF around 2.9 bohr which impedes a straightforward determination of the equilibrium distance (see below). Another large irregularity appears around 3.4 bohr.
Again the smooth BSE@COHSEX@HF curve is closer to that obtained within CC3 than the BSE@{\GOWO}@HF curve, although the differences are small.
Similar to the LiF results obtained above for the small cc-pVDZ basis, we observe again that including self-consistency in the COHSEX calculation lowers the total energy, thereby worsening the agreement with the coupled-cluster reference data.
Finally, we note that for LiF the BSE@HF total energies are slightly closer to the reference CC3 results than those obtained within BSE@COHSEX@HF.
However, as we will see in the following, the PES obtained within BSE@COHSEX@HF results are, in general, better than those obtained within BSE@HF.

The PES of all diatomic molecules, except the smallest two (H$_2$ and LiH), show similar trends as LiF, i.e., small differences between the BSE@COHSEX@HF and BSE@{\GOWO}@HF total energies and a relatively large difference with respect to the BSE@scCOHSEX total energies.
Therefore, we conclude that the self-consistency has a much larger influence on the PES than the difference in the COHSEX and $GW$ self-energies.

The PES of the HCl, N$_2$, CO and BF molecules obtained within BSE@{\GOWO}@HF all exhibit small irregularities, while those in F$_2$ are very large, preventing a simple determination of the F$_2$ equilibrium distance (see below).
Again, BSE@COHSEX@HF is in excellent agreement with the CC3 results and even slightly better than those obtained within BSE@{\GOWO}@HF,
and, most importantly, the PES obtained within BSE@COHSEX@HF (and BSE@scCOHSEX) are devoid of irregularities and discontinuities.

In Table \ref{Table:Req} we report the equilibrium distances obtained within the various BSE approaches and we compare them to the CC3 reference values and to experiment. As mentioned before, the irregularities in the PES can prevent a straightforward determination of the equilibrium distance. Therefore, following Ref.~\citenum{Loos_2020}, for LiF and F$_2$ a Morse potential was used to fit the total energies in order to estimate the equilibrium distance.
Although the total energies obtained within BSE@scCOHSEX were not as accurate as those obtained using perturbative QP energies, adding self-consistency to the COHSEX approach improves the equilibrium distances. In summary, while BSE@COHSEX@HF yields the smallest errors for the total energies, BSE@scCOHSEX yields the smallest errors for the equilibrium distances.
\begin{table}[t]
\caption{Equilibrium distances (in bohr) obtained in the cc-pVQZ basis set.
The experimental values are extracted from Ref.~\citenum{HerzbergBook}. 
The results in brackets for LiF and F$_2$ were obtained by fitting the total energies to a Morse potential since the irregularities in the PES precluded a direct evaluation.}.
\begin{center}
\begin{tabular}{lcccccccc}
\hline
\hline
& H$_2$ & LiH & LiF &  HCl & N$_2$ & CO & BF & F$_2$ \\
\hline \\
CC3 & 1.402 & 3.019 & 2.963 & 2.403 & 2.075 & 2.136 & 2.390 & 2.663     \\
BSE@HF & 1.402 & 3.014  & 2.954 & 2.400 & 2.065  & 2.120 & 2.378 &  2.631    \\
BSE@G$_0$W$_0$@HF  & 1.399 & 3.017  & (2.973) & 2.400 & 2.065 & 2.134 & 2.385 & (2.638)   \\
BSE@COHSEX@HF  & 1.399 & 3.014  & 2.961 & 2.400 & 2.066 & 2.125 & 2.379 & 2.635   \\
BSE@scCOHSEX  & 1.401 & 3.016 & 2.963  & 2.404 & 2.070 & 2.130 & 2.387 & 2.650 \\
Experiment & 1.401	& 3.015 & 2.948 & 2.409 & 2.074 & 2.132 & 2.386 & 2.668 \\
\hline\hline
\end{tabular}
\end{center}
\label{Table:Req}
\end{table}

Finally, in order to estimate the influence of the QP energies on the BSE total energies, we report the ionization potentials (IP) and the HOMO-LUMO gaps at the equilibrium distance corresponding to each level of theory for the various BSE approaches in Tables \ref{Table:IP} and \ref{Table:gaps}, respectively, and we compare to experimental data (when available).
For the IP we also report the CCSD(T)/def2TZVPP data of Ref.~\citenum{Krause_2015} which are in good agreement with the experimental values with the exception of H$_2$.
Comparing the differences in the IP with the differences in the PES, there does not emerge a clear link between the two.
Although the IP obtained within COHSEX@HF and {\GOWO}@HF show large differences, the differences between the corresponding BSE total energies are small.
Instead, the differences in the IP between scCOHSEX and COHSEX@HF are small (except for N$_2$) but the differences in the corresponding total energies are large.
Similarly, the differences in the IP between HF and {\GOWO}@HF are small but the differences in the corresponding total energies are large.
An equivalent analysis holds for the HOMO-LUMO gaps.
Moreover, despite the fact that COHSEX@HF yields IP and HOMO-LUMO gaps significantly worse than those obtained within {\GOWO}@HF when compared to the experimental values, the corresponding BSE total energies are very similar (except for the irregularities in {\GOWO}@HF@BSE). Therefore, at least for the small molecules discussed here, the BSE total energies obtained within ACFDT seem to be robust with respect to the screening used in the calculation of the the underlying QP energies.
Instead, the total energies are sensitive to the screening that enters the BSE.
Within BSE@COHSEX@HF and BSE@{\GOWO}@HF this quantity is identical since in both cases it is calculated from the HF orbitals and energies.
However, when one includes self-consistency, the screening changes and it has a significant influence on the total energy.
We can therefore conclude that the screened Coulomb potential is the key quantity in the calculation of correlation energies within the ACFDT@BSE formalism, and ultimately dictates the accuracy of the total energy.
Nevertheless, when screening is completely neglected in the calculation of the QP energies, e.g., in BSE@HF, this also has a significant influence on the results.

Finally, we note that, although BSE@COHSEX@HF gives the best PES and BSE@scCOHSEX the best equilibrium distances, the COHSEX@HF and scCOHSEX ionization potentials and HOMO-LUMO gaps are not very good. Therefore, it would be desirable to find an improved static self-energy which could give both smooth PES and good QP energies.
\begin{table}[t]
\caption{Ionization potentials (in eV) at the equilibrium distance obtained in the cc-pVQZ basis set except for the CCSD(T) values from Ref.~\citenum{Krause_2015} which have been obtained in the def2-TZVPP basis. The experimental values are extracted from Ref.~\citenum{vanSetten_2015}}
\begin{center}
\begin{tabular}{lcccccccc}
\hline
\hline
& H$_2$ & LiH & LiF &  HCl & N$_2$ & CO & BF & F$_2$ \\
\hline \\
HF  & 16.17 & 7.99 & 12.94 & 12.98 & 16.76 & 15.07 & 11.01 & 18.02   \\ 
G$_0$W$_0$@HF  & 16.57 & 8.26 & 11.59 & 12.98 & 17.33 & 14.91 & 11.41 & 16.50  \\
COHSEX@HF      & 18.05 & 9.52 & 13.82 & 14.49 & 19.48 & 16.69 & 12.86 & 18.88  \\
scCOHSEX       & 17.83 & 9.21 & 13.12 & 14.02 & 17.52 & 15.79 & 12.45 & 18.00  \\
CCSD(T)        & 16.40 & 7.96 & 11.32 & 12.59 & 15.57 & 14.21 & 11.09 & 15.71  \\
Experiment     & 15.43 & 7.90 & 11.30 & 12.79 & 15.58 & 14.01 & 11.00 & 15.70  \\
\hline\hline
\end{tabular}
\end{center}
\label{Table:IP}
\end{table} 
\begin{table}[t]
\caption{HOMO-LUMO gaps (in eV) at the equilibrium distance obtained in the cc-pVQZ basis set. The experimental values are extracted from Ref.~\citenum{Maggio_2016}}
\begin{center}
\begin{tabular}{lcccccccc}
\hline
\hline
& H$_2$ & LiH & LiF &  HCl & N$_2$ & CO & BF & F$_2$ \\
\hline \\
HF  & 20.08 & 8.08 & 12.72 & 15.76 & 20.95 & 18.55 & 13.15 & 20.80   \\ 
G$_0$W$_0$@HF  & 20.24 & 8.04  & 11.31 & 15.20 & 20.24 & 17.33 & 12.90 & 17.32   \\
COHSEX@HF  & 21.59 & 9.27 & 13.54 & 16.45 & 21.38 & 18.44 & 13.97 & 18.14  \\
scCOHSEX  & 21.57  & 8.99 & 12.84 & 16.07 & 20.09 & 17.93 & 13.73 & 17.81 \\
Experiment & & 8.24 &  & &  & &  & 16.94 \\
\hline\hline
\end{tabular}
\end{center}
\label{Table:gaps}
\end{table} 
%
\section{Conclusions}
\label{Conclusions}

We have demonstrated that COHSEX is a promising approach to obtain quasi-particle energies for the calculation of potential energy surfaces.
Contrary to {\GOWO} and partially self-consistent $GW$ approaches, COHSEX yields results without irregularities and discontinuities.
We have illustrated this feature by calculating the ground-state potential energy surfaces of diatomic molecules.
Moreover, we have shown that BSE total energies of diatomic molecules using COHSEX quasi-particle energies obtained perturbatively on top of a Hartree-Fock calculation are in good agreement with accurate coupled-cluster results. Finally, we showed that including self-consistency in the COHSEX approach for both quasi-particle energies and orbitals, in order to make the results independent of the starting point, worsens the total energies but improves the equilibrium distances.
This is mainly due to variations in the screening $W$ that enters the BSE.

 
\begin{acknowledgement}
JAB and PR thank the French Agence Nationale de la Recherche (ANR) for financial support (Grant agreements ANR-18-CE30-0025 and ANR-19-CE30-0011).
PFL thanks the European Research Council (ERC) under the European Union's Horizon 2020 research and innovation programme (Grant agreement No. 863481) for financial support.
This study has also been partially supported through the EUR grant NanoX n$^\text{o}$ ANR-17-EURE-0009 in the framework of the ''Programme des Investissements d'Avenir''.
\end{acknowledgement}

\providecommand{\latin}[1]{#1}
\providecommand*\mcitethebibliography{\thebibliography}
\csname @ifundefined\endcsname{endmcitethebibliography}
  {\let\endmcitethebibliography\endthebibliography}{}


\end{document}